\documentclass[final]{isss7}

\usepackage{graphicx}
\usepackage{times}

\title{Tutorial introduction to Virtual Reality:\\
What possibility are offered to our field?}
\author{Akira Kageyama and Nobuaki Ohno}
\affiliation{
Earth Simulator Center,\\
Japan Agency for Marine-Earth Science and Technology,\\
3173-25 Showa-machi, Yokohama 236-0001, Japan
}

\abstract{
The virtual reality (VR) provides us a three-dimensional,
immersive, and fully interactive visualization environment.
To make the best use of the VR's potential in scientific visualization,
a VR visualization software named VFIVE 
has been developed for the CAVE-type VR system.
VFIVE enables simulation researchers to analyze three-dimensional
scalar and vector fields by various visualization 
methods including real time volume rendering
in the CAVE's room sized booth.
Some basic visualization tools of VTK have been integrated to VFIVE, too.
}

\begin{document}
\maketitle

\section{Introduction}

The possibility offered by Virtual Reality (VR) to the simulation science
is an innovation of the scientific visualization.
VR visualization is not a theory, but rather already a practical technology 
that can be used in our daily research routine.

Simulation science is supported by two key technologies;
computation and visualization.
The technology of computation, or computer, is still keeping an exponential growth.
In this sense, we are living in a golden age of simulation science.
However, a high-rise building should be supported by balanced pillars.
Unfortunately, the development of visualization technology does not 
catch up with its counterpart.
It would be absurd if we had to spend a month to understand 
the simulation data obtained by one-hour-job on ultra super computer.

In the Stone Age of computer simulation 
when 1-dimensional (1-D) phenomena are mostly simulated,
simple graph plot on a piece of paper would have been 
sufficient to understand or to visualize the data.
When 2-D simulation were common, iso-line plots and contour color plots
on papers or computer monitor screens
would have been enough.

In the beginning of 1990's,
3-D simulations with the grid size of $O(100^3)$ was not rare.
We familiarized ourselves with newly introduced 
visualization technology at that time;
it was so called ``visualization software'' (such as AVS) on 
graphic workstations (GWS).
We could instantly see an isosurface of $100^3$ data points through
the GWS's monitor.
we could zoom, rotate, color the objects, and
change the isosurface levels, via graphical user interface using
the mouse.
By the interactive manipulation through the GWS's monitor,
we could grasp 3-D structure of the numerical data.
This visualization technology based on GWS was certainly powerful enough
at that time.

Now, the complexity of target phenomena of 
the present high-end super computer,
such as the Earth Simulator\footnote{
http://www.es.jamstec.go.jp/esc/eng/index.html
},
are extremely high;
the typical grid size is $O(1000^3)$.
Even if we could make, render, and rotate an isosurface object
on the $O(1000^3)$ grid points in real time on the GWS or PC 
with the advanced graphics card,
the complicated shape itself
of the isosurface prevents us 
from the intuitive and instant understanding of the 3-D 
structure of the scalar field.

Another limitation of the GWS-based visualization technology
is the ability (or inability) of visualization of vector fields.
In our research fields, 
fluid flows, electric fields, and magnetic fields are all 3-D vector fields.
We need to grasp spatial structure of the 3-D vector fields defined on
$O(1000^3)$ grid points in order to understand what was 
simulated in the simulation.
This is really a challenging task and obviously beyond the outdated,
GWS-based visualization technology.
Simulation scientists really need an 
innovation in the visualization technology that suits
the modern high-end super computer.
And we believe that the virtual reality (VR) is the answer.

Everyone who experiences a modern VR system for the first time
would be surprised by its high quality of mimicked reality.
They feel like they are deeply absorbed, or really standing, in the mimicked world.
In order to produce such a deep absorption into the VR world,
there are three important visual factors\footnote{
Sonifications~[Tamura:2001]
and haptics are other important VR technology that could be
applied in scientific visualization in future.
};
(1) stereo view, (2) immersive view, and (3) interactive view.
Among various kinds of available VR hardwares,
we believe that the CAVE~[Cruz-Neira:1993] system suits best to our purpose of the
visualization of large scale 3-D scalar and vector fields.
The CAVE is developed at Electric Visualization
Laboratory (EVL), University of Illinois\footnote{
http://www.evl.uic.edu/info/index.php3}.
We installed our CAVE system at Earth Simulator Center,
Japan Agency for Marine-Earth Science and Technology (JAMSTEC)
in 2002, and named it BRAVE.

\section{Hardware of VR Visualization: BRAVE}
\begin{figure}[ht]
\centerline{\includegraphics[width=0.99\linewidth]{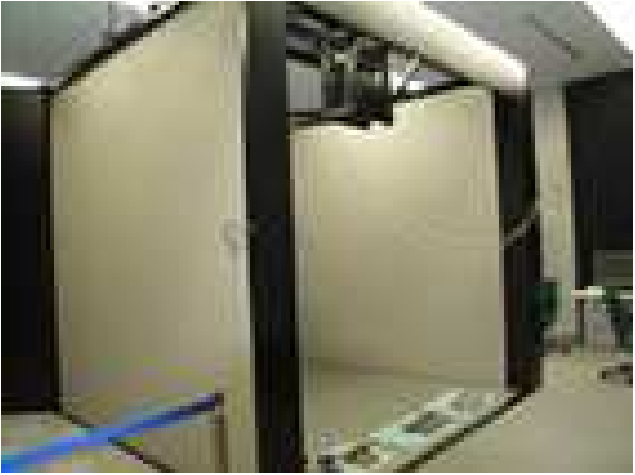}}
\caption{\label{fig:brave}
The CAVE system in the Earth Simulator Center, called BRAVE.}
\end{figure}

BRAVE is a cubic room with the size of 
$3\hbox{m}\times 3\hbox{m}\times 3\hbox{m}$.
It has four stereo screens (three walls and a floor).
See Fig.~\ref{fig:brave}.
The viewer stand in the BRAVE room on the floor screen,
wearing stereo glasses,
with a portable controller called wand.
Stereo images are projected on the three walls and the floor by four projectors.
Since the viewer inside the BRAVE room is 
surrounded by stereo images on the three walls and the floor,
he or she can look around like a owl in the BRAVE room,
keeping the wide range of stereo view angle.
This strongly enhances the immersive sense of the viewer.
The refresh rate of the image is 96 Hz:
The right-eye image and the left-eye image are 
projected alternately with 48 Hz each.
The image refresh is synchronized with the stereo glasses
by the infrared.
The images on the boundaries between the walls and the floor are smoothly
connected.
The viewer can easily forget the existence of the boundaries.
Actually, some people had banged on the walls of the BRAVE.

BRAVE has a magnetic tracking system 
to detect the position and direction of the viewer's eyes
(by a sensor on the stereo glasses),
and the viewer's hand (by another sensor on the wand).
All the projected images are automatically adjusted in real time
following the viewer's head motion.
Therefore, everything looks natural from the viewer
who can tilt, walk, sit, or even jump in the BRAVE room
to observe 3-D objects in the virtual world.

The wand, which has 3 buttons and 1 joystick,
is used to the interface with the virtual world (or simulation data).
For example, when viewer presses a wand button, a virtual menu panel 
appears in front and he or she can choose a visualization method
by shooting a menu box by a virtual laser beam emitting from the wand.
If you presses another button, a virtual tracer particle appears
at the tip of the wand,
when he or she has chosen the particle tracer menu,
and when the button is released,
the particle ``flies'' under the eyes 
following the velocity field showing the flow structure
by its trajectory.
This kind of user interface of the wand is programmable with
commercially available basic VR library called CAVE lib.

The OpenGL is used to model the virtual world;
it is used to define the visualization objects and its illumination.
A BRAVE application program, or more generally, a CAVE application program,
is essentially a kind of common computer graphics (CG) software
without the projection part which is automatically processed by the CAVE lib.
So, if you are familiar with CG programming using OpenGL, 
it would be easy for you to make your own CAVE application.

\section{Software of VR Visualization: VFIVE}
\begin{figure}[ht]
\centerline{\includegraphics[width=0.99\linewidth]{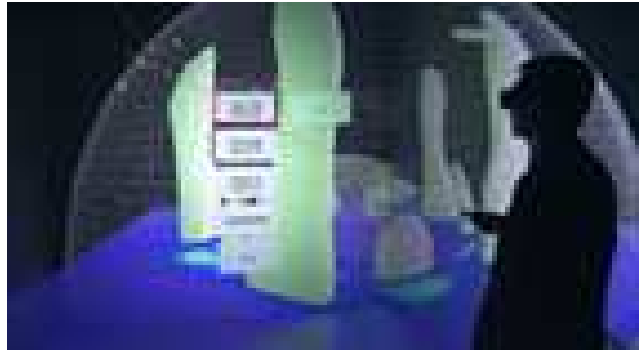}}
\caption{\label{fig:v5menu}Interactive menu of our original
VR visualization software VFIVE.}
\end{figure}

Since most of simulation researchers are not familiar with
CG programming, they feel it difficult to make their own
visualization softwares for the CAVE.
In order to help them, we have been developing a
general purpose VR visualization software named VFIVE
for the scientific VR visualization in the CAVE.
The development of VFIVE started around 1999~[Kageyama:2000]
and the latest stable version is v3.7.
The present development version with major revise is v3.8.
The very final version of VFIVE will be v5.

\subsection{VFIVE v3.7}
The basic design and core functions
of VFIVE are almost fixed by v3.7,
which are summarized in Table~\ref{table:v3.7}.
A great emphasis was placed on fully interactive control of the
visualization methods, data, and the user interface 
in the CAVE's VR environment.
For example,
one can control the {\sl isosurface} levels by vertical motion of the hand with the wand;
it is also possible to change the slice position of {\sl orthoslicer} by the wand's motion.
From our experience of using the CAVE as a scientific visualization tool
from the end of 1990's~[Kageyama:1999], 
we see that a common pitfall is to use it 
just as a (very expensive) ``3-D object viewer''.
We should make the best use of CAVE's VR features;
that is stereo, immersive, and especially interactive view.

We implemented several kinds of visualization methods for vector fields.
All of them make the best use the VR features of the CAVE.
For example, {\sl vector arrows} shows tens of small arrows
around your hand, within an invisible sphere of diameter of 2 feet.
The direction and length of each arrow show the vector at the point.
As you move your hand, all the arrows follow your hand's motion,
changing the direction and length 
as the sampled point of each arrow changes in the vector field.
The interpolation is automatically applied in real time.

\begin{table}[ht]
\renewcommand{\arraystretch}{1.2}
\vspace{-.3cm}
\caption{VFIVE v3.7}\label{table:v3.7}
\vspace{-.5cm}
\begin{center}
\begin{tabular}{|c|c|c|} \hline
                       & function  & Figure \\ \hline\hline
user interface & virtual menu panel & fig.~\ref{fig:v5menu}\\ \hline
                       & virtual laser beam  & fig.~\ref{fig:v5menu}\\ \hline
scalar field vis. & ortho-slice planes & fig.~\ref{fig:isosurf} \\ \hline
                                      & local slice plane           & not shown \\ \hline
                                      & isosurface                    & fig~\ref{fig:isosurf} \\ \hline\hline
vector field vis. & tracer particles   & fig.~\ref{fig:tracers} \\ \hline
                                       & field (force) lines & not shown \\ \hline
                                       & vector arrows     & fig.~\ref{fig:arrows} \\ \hline
                                       & spotlighted particles  & now shown\\ \hline
\end{tabular}
\end{center}
\end{table}

\subsection{VFIVE v3.8}
Recently, we have improved the VFIVE in two aspects:
First, we included an important scalar visualization method
that was missing in the previous version; volume rendering~[Drebin:1988]. 
Another improvement is that we have integrated some basic modules of
VTK\footnote{
http://public.kitware.com/VTK/
} into VFIVE.

The volume rendering is a
visualization method of scalar fields
that is useful, for example, to show the distribution of plasma density $\rho(x,y,z)$
of the 3-D global simulation of the magnetosphere.
It is known that $\rho$ is highly localized near the Earth.
The isosurface is not useful for the visualization of this kind of scalar field;
you need to try nearly infinite number of different isosurface levels to
grasp the overall 3-D distribution of $\rho$.
Similarly, the orthoslicer is not useful because one need to try
a lot of slices to grasp 3-D distribution of $\rho$.
The volume rendering fits best to this kind of 3-D scalar field.
We have developed a volume rendering module of VFIVE and 
slotted into v3.8.

Generally, the volume rendering is considered as
a ``heavy'' visualization technique for the CAVE-type VR systems
since it requires millions of ray cast calculation per second to
realize the real time response to viewer's eyes motion in the CAVE room.
We succeeded in realizing a very fast volume rendering
by the 3-D texture-map technique~[Schroeder:2002]. 
In this technique, many semi-transparent 
texture mapped slices,
which are orthogonal to the viewer's line of sight, 
are pulled out from the a 3-D volumetric data.
The 3-D volumetric data is made in advance 
from the scalar field (such as $\rho$) 
by specifying the colors and opacities of each box element (voxel).
The texture slices are blended by the graphics card.
Therefore, we can avoid the heavy calculations of software ray casting.
This is the reason why we can carry out the fast volume rendering in the CAVE.
For example, the refresh rate of the volume rendering shown in Fig.~\ref{fig:volrend}
(vorticity distribution of Earth's outer core)
is several frames per second, which is satisfactory for our visualization purpose.

\begin{figure}[ht]
\begin{center}
\includegraphics[width=0.99\linewidth]{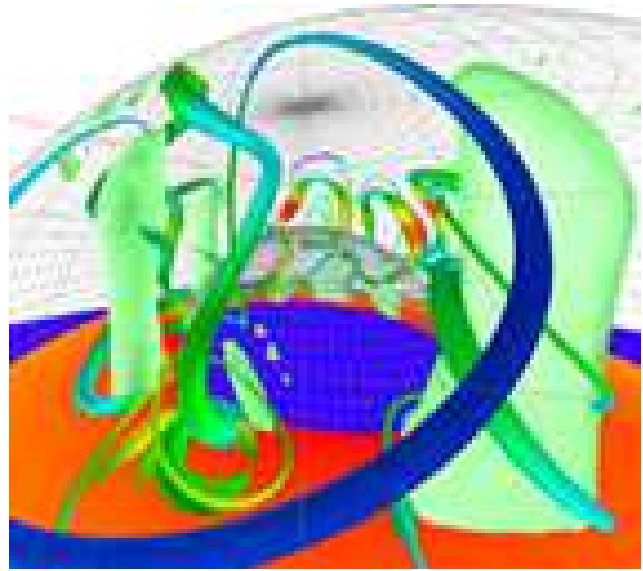}
\end{center}
\caption{\label{fig:tube}Stream tube by VTK in VFIVE v3.8.
The tube's diameter and color changes according to local amplitude of
the vector field.
}
\end{figure}

\begin{figure}[ht]
\centerline{\includegraphics[width=0.99\linewidth]{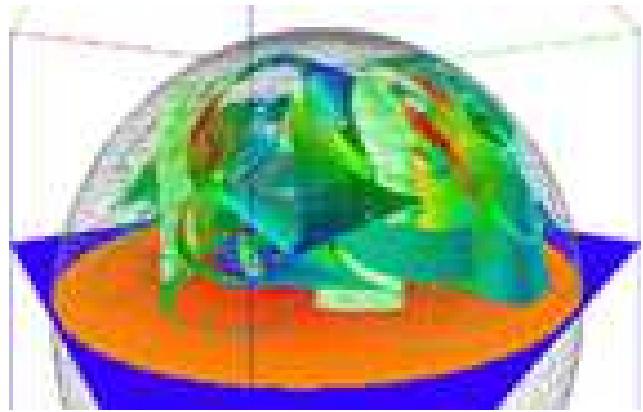}}
\caption{\label{fig:ribbons}Stream surfaces by VTK in VFIVE v3.8.
The color changes according to the local amplitude of the vector field.}
\end{figure}

The VTK (Visualization Tool Kit) is an general visualization software
that includes many visualization methods, from basic ones
to sophisticated ones, with open source codes.
It was an attractive idea for us to combine VFIVE's interactive
visualization environment in the CAVE with
VTK's sophisticated visualization gismos.
Recently, we have succeeded to integrate some basic visualization methods
of VTK into VFIVE v3.8.
For instance, we can show VTK's iso-contours,
tubed flow lines (Fig.~\ref{fig:tube}), 
stream surfaces (Fig.~\ref{fig:ribbons}) in the BRAVE (see Table\ref{table:v3.8}).

\begin{table}[ht]
\renewcommand{\arraystretch}{1.2}
\vspace{-.3cm}
\caption{New features in VFIVE v3.8}\label{table:v3.8}
\vspace{-.5cm}
\begin{center}
\begin{tabular}{|c|c|c|} \hline
                       & function  & Figure \\ \hline\hline
scalar field vis. & volume rendering & fig.~\ref{fig:volrend} \\ \hline
                         & isocontours (VTK)   & not shown \\ \hline	
vector field vis. & stream lines (VTK)   & fig.~\ref{fig:tube} \\ \hline
                        & stream surfaces (VTK) & fig.~\ref{fig:ribbons} \\ \hline
\end{tabular}
\end{center}
\end{table}

\begin{figure}[ht]
\centerline{\includegraphics[width=0.99\linewidth]{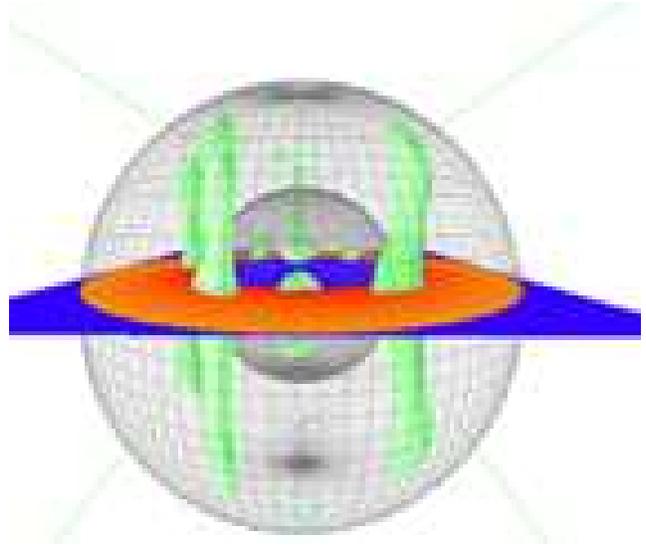}}
\caption{\label{fig:isosurf}Scalar field visualization by isosurface,
and orthoslicer.}
\end{figure}

\begin{figure}[ht]
\centerline{\includegraphics[width=0.99\linewidth]{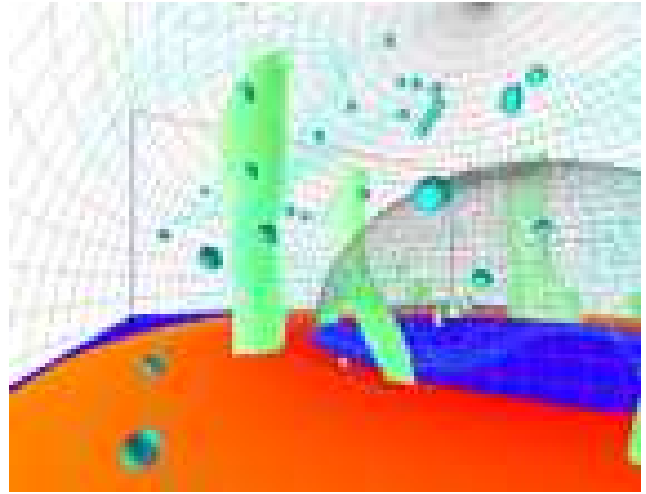}}
\caption{\label{fig:tracers}Vector field visualization by tracer particles.
When you press a wand button in the CAVE room, a tracer particle appears
at the tip of the wand.
You can intuitively control the starting position of the 3-D vector field
by the wand motion.}
\end{figure}

\begin{figure}[ht]
\centerline{\includegraphics[width=0.99\linewidth]{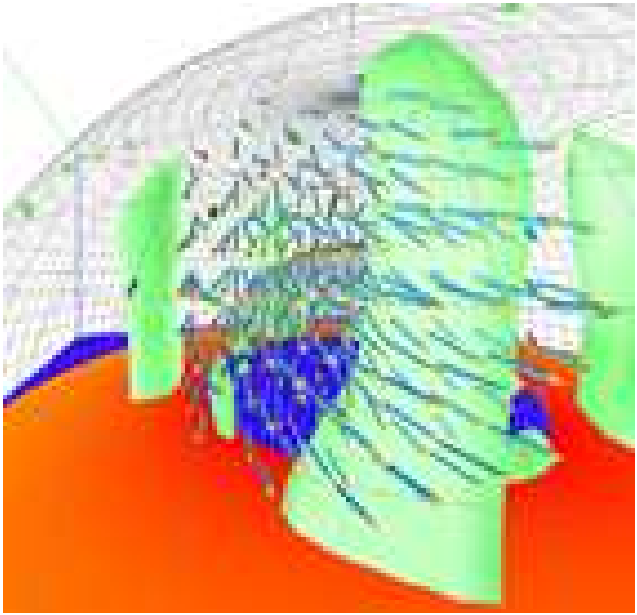}}
\caption{\label{fig:arrows}Interactive vector field visualization by {\sl arrows} with
the isosurface rendering.
The arrows change the length and direction,
following your hand's motion in real time
with the automatic interpolation of the vector field.
}
\end{figure}

\begin{figure}[ht]
\centerline{\includegraphics[width=0.99\linewidth]{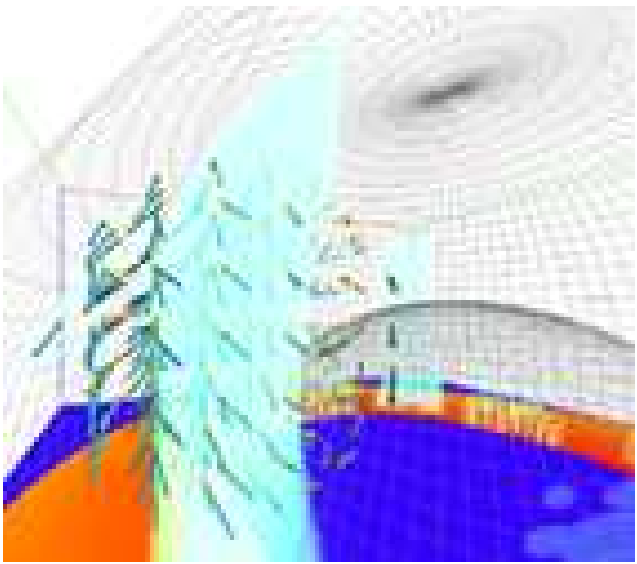}}
\caption{\label{fig:volrend}Real time volume rendering implemented in VFIVE v3.8 with 
arrows. Compare with isosurface rendering in Fig.~\ref{fig:arrows}.
The high speed volume rendering is realized by the 3-D texture-map technique.
}
\end{figure}

\section{Summary}
There has been an serious imbalance between the computer technology
and visualization technology.
Ultra high performance of future computer would become futile existence
if it were not for the symmetrical performance of the data visualization.
The advanced visualization should be performed in
a 3-D, interactive, and immersive environment,
and it can be offered by the virtual reality (VR) technology.

We speculate that every super computer center in the world
will install the CAVE-type VR facility for
the VR visualization, and use the software like VFIVE 
for practical and productive visualization routine.

\acknowledgments{
We thank all the other members of
Advanced Perception Research Group of Earth Simulator Center
for their help in the development VFIVE v3.8.
We also thank Prof.~Tetsuya Sato, the director-general
of Earth Simulator Center,
for fruitful discussion and encouragement.
}


\end{document}